# Transport Triggered Array Processor for Vision Applications

Mehdi Safarpour, Ilkka Hautala, Miguel Bordallo Lopez, and Olli Silven

Center for Machine Vision and Signal Analysis, University of Oulu, Finland [1]
mehdi.safarpour [at] oulu.fi

**Abstract.** Low-level sensory data processing in many Internet-of-Things (IoT) devices pursue energy efficiency by utilizing sleep modes or slowing the clocking to the minimum. To curb the share of stand-by power dissipation in those designs, near-threshold/sub-threshold operational points or ultra-low-leakage processes in fabrication are employed. Those limit the clocking rates significantly, reducing the computing throughputs of individual processing cores. In this contribution we explore compensating for the performance loss of operating in near-threshold region ($V_{dd}$ =0.6V) through massive parallelization. Benefits of near-threshold operation and massive parallelism are optimum energy consumption per instruction operation and minimized memory roundtrips, respectively. The Processing Elements (PE) of the design are based on Transport Triggered Architecture. The fine grained programmable parallel solution allows for fast and efficient computation of learnable low-level features (e.g. local binary descriptors and convolutions). Other operations, including Max-pooling have also been implemented. The programmable design achieves excellent energy efficiency for Local Binary Patterns computations.

## 1 Introduction

With the decreasing costs of cameras and wireless communications, an unprecedented growth in the number of imaging sensors deployed in our environment is taking place. This is coupled to the growth of Internet of Things (IoT) and cloud computing that transforms the little-data from distributed sensors to centralized big-data. Examples of rapidly growing applications include Advanced Driver Assistance Systems (ADAS), data gathering using drones, surveillance systems and service robotics. These applications try to interact with the environment or to extract information from the scene, necessitating high performance computing, while demanding extreme energy efficiency if they depend on energy harvesting or battery power.

[1]The support of Academy of Finland for project ICONICAL (grant 313467) and 6Genesis Flagship (grant 318927) is gratefully acknowledged.

In conventional embedded processors, up to 70% of the power dissipation is due to the instruction and data supply [2] making those the prime targets for architectural optimization. On the other hand, in low level computer vision most of the operations deal with neighborhoods of pixels, providing opportunities to avoid memory round trips in local processing. This calls for application specific architectures [13], and has lead to array processor proposals, mostly in a 2-D mesh configuration [7]. Unfortunately, they seldom provide for flexible programmability, and as such mostly serve as energy efficiency and raw throughput benchmarks.

Previous studies have demonstrated the usefulness of GPUs and 1-D SIMD processors for low-level vision operations [12]. Although these architectures tend to suffer from memory and I/O bottlenecks due to frequent data transfer to and out of the PEs [24], several studies [7,3,24,21] have demonstrated their attractiveness.

The sizes of the reported massive processing arrays have varied, e.g., from 170×120 [7] to 256×256 PEs [5], while both digital and mixed mode technologies have been employed. All of these works are very similar in implementation. It has been shown that analog/mixed signal based massive arrays possess superior area-energy efficiency, but the analog computation is susceptible to noise in deep sub-micron technology. This issue is almost non-existent for the digital counterpart [4,21].

TTA cores were adopted as processing element of the presented array, due to simplicity in design and availability of a design tool-chain. In previous works single core and coarse-grained high performance TTA based solutions were already developed and demonstrated. Ijzerman et al. [12] proposed programmable SIMD TTA-based accelerator for convolutional neural networks. Also, in [11] a coarse-grained multi-core TTA was designed for video coding applications.

In the current contribution, we address the design of a massive array processor using the TTA architecture template. To the best knowledge of the authors, this is the first such study. For the design, we used the available advanced TTA co-design environment [8]. The motivation for the study stems from the observation of potential energy efficiency benefits attainable from ultra-low-leakage silicon technologies and operating in near-threshold region. However, this approach is penalized by exponential increase in circuit delay. The massive parallelism offsets the speed penalty from the low clock frequency, consequently, we decided to realize the design using a near-threshold technology [6].

The clocking frequency is not a constraint in massive arrays used for most vision applications (frame intervals are long enough to finish a large sequence of image operations), so one extreme design approach is to operate in sub-threshold regime with optimum sub-threshold voltage that minimizes the energy per instruction. We notice the energy efficiency of sub-threshold voltage designs (e.g. sub-threshold voltage FFT processor [22] with $155nJ$ per 16-b 1024-point FFT, clocked at only 10 kHz in $180nm$ technology). Unfortunately, our tool chain didn't allow comparable experiments. However, near-threshold design space is explored in this work.

In addition, to show the advantages of the programmability of our architecture, we evaluate it with relevant low-level image processing operations, including learnable local descriptors, variable convolutions and Max-pooling operation. The operations are components in the inference stage of the current state-of-the art computer vision algorithms. In all the operations our architecture shows its advantages in memory bound algorithms since it does not need to flush data back and forth between memories [18].

## 2    System architecture

### 2.1    The array processor architecture

In our proposed architecture, all PEs are directly connected to neighboring PEs. The instruction memories are shared between groups of processors. Vertical and horizontal indices are assigned to each PE to make it feasible to selectively run instructions or to form PE groups, where each group executes its own instruction stream. As an example, an 8 × 8 example architecture is shown in Fig. 1. As depicted in the figure, each PE is connected to the neighborhood register bank that contains its immediate neighbors in eight directions.

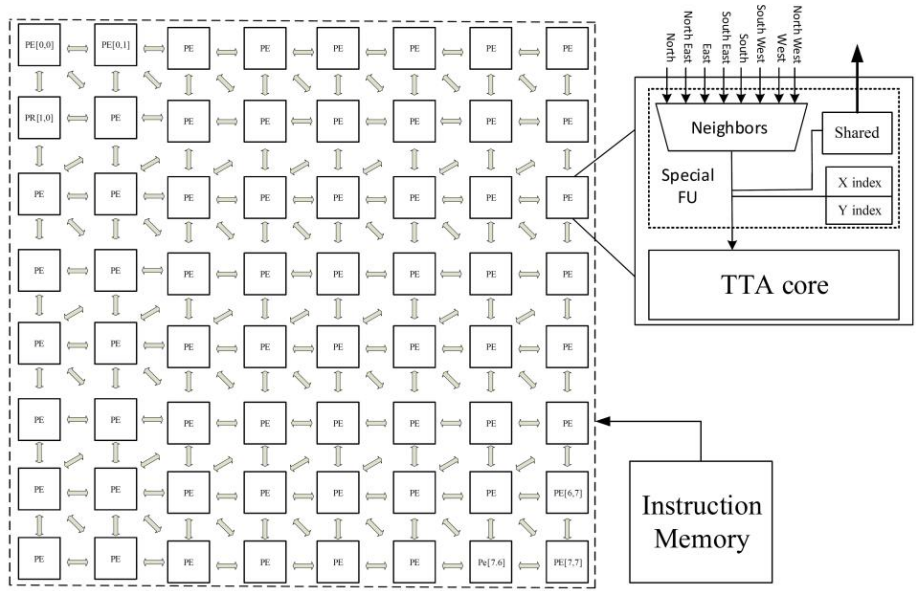

**Fig.1.** General view of the massive array architecture

Some image processing operations require activating only a small set of PEs, while the rest could function as memory. For example, in some forms of maxpooling non-overlapping windows must be selected. For the purpose of grouping a bundle of PEs to certain operations, each PE is aware of its horizontal and vertical index. This enables instructing a PE to remain idle depending on its indices.

The PEs where chosen to be based on the TTA architecture due to their relative architectural simplicity, ease of design and the exposed bypass network of the processors. In this architecture, similar to a general approach in massive array processors, all processing elements receive a single instruction stream and simultaneously execute the same instruction on their local data [4]. However, in our scheme multiple instruction memories can feed different groups of PEs and each PE can multiplex between different instruction memories.

Vision applications usually require a large number of computations, especially for pixel level operations. Generally, the frame rate for the cameras integrated to current embedded systems do not exceed 120 frames per second, while the rates typically range from 30 to 60 Hz. Even applications, such as visual odometry that usually require high frame rate, rarely exceed rates higher than a few hundred frames per second.

In this context, we aimed at an architecture that could flexibly employ varying numbers of processors (e.g., PE arrays from $3 \times 3$ to $128 \times 128$), while we could operate them at a very low frequency and voltage, using ultra-low power strategies. Moreover, the array can be put in sleep mode during frame intervals, essentially functioning in a race-to-sleep mode, which significantly reduces the average power consumption [1].

## 2.2    Sensor Processor Arrangement

Generally, two forms of arrangement can be considered for 2-Dimensional sensor and digital processing arrays [24]. In the first one, each pixel is coupled to a pixel level Analog-to-Digital Converter (ADC) [15] and a PE and the ADC directly writes into the corresponding PE. This approach mostly is used in applications where the number of sensors is limited. In the second approach, sensor plane and processing array are separated.

Two examples of this approach are shown in Fig. 2 and Fig. 3. In Fig. 2, row parallel ADCs quantize image pixels column by column and fed the output into first column of the processing array (alternatively a single ADC can be coupled to a 1-D column buffer and the buffer is flushed into the array) [15]. Subsequently, data is propagated in the array in a wave manner. This way the maximum number of cycles to load a totally new image onto the array is equal to the number of columns. In case that the processing array is not large enough to accommodate the whole image, a moving window called *Fovea arrangement* (Fig. 3) [24], swept throughout the image plane, is read and fed into the array. Benefiting from the exposed bypass networks of TTA, our design provides means to pass data from PE to PE efficiently without any extra hardware.

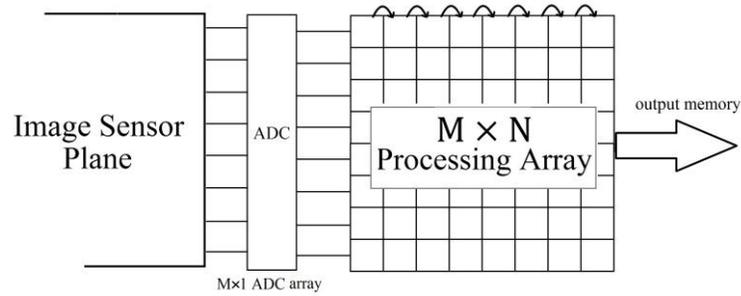

**Fig.2.** Sensor readout in column by column fashion

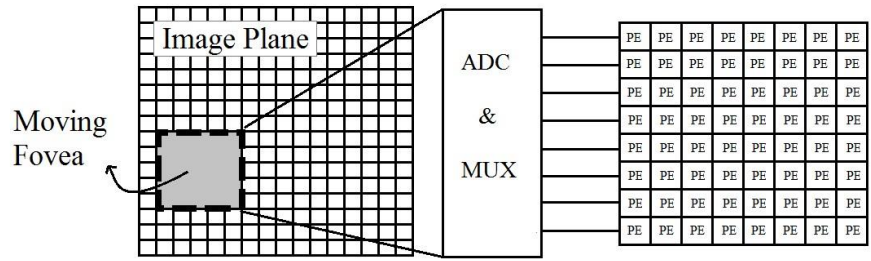

**Fig.3.** Fovea arrangement where number of processor is less than number of sensors

## 3 System implementation

The proposed architecture was implemented and simulated using SystemC, since the TCE toolchain provides means to integrate its cycle accurate simulator with custom designs that are implemented this way. At first, after experimenting with basic image processing operations, the TTA core was pruned to consume minimum energy and logic gates. The detail of the TTA core employed as the architecture PE is presented in Table 1, while a scheme depicting the core itself is shown in Fig. 4.

**Table 1.** Detail of general TTA core

| Component | Details | Quantity |
|---|---|---|
| ALU | ADD, EQ, GTU | 1 |
| Logic | AND, OR, XOR, SHR, SHL | 1 |
| Special custom FU | Neighborhood Shared register, Inputs ports | 1 |
| Register file | 16-Bit Registers for temporary data storage | 4 |
| Boolean Register file | For storage outcome of logic operations | 2 |
| Instruction memory width | 23 b | |
| Short Immediate | 16 b | |

### 3.1   Neighbor Communication Functional Unit

A special functional unit (FU) to communicate with the eight adjacent neighbors of each PE was designed using behavioral models written in SystemC and VHDL. This unit contains both the vertical and horizontal indices of the PE. It consists of two internal ports (one input and one output) that are connected to the main bus of the TTA core and nine external ports (eight inputs and one output) that communicate with the neighbouring PEs.

Each PE can store its output on a register named *Shared register* which can be read only by the neighbouring PEs. One of the external ports of the custom functional units is devoted to this register. The other eight external input ports named *North, North East, East, South East, South, South _West, West, North _Wes*t, are connected to corresponded Shared register of neighbouring PEs. The custom functional units provide three instructions *read neighbour*, *read _index* and *write _Shared*. Example transports to read and write from the custom FU are shown in Table 2. The SystemC model of the FU was integrated to the TTA based PE core. A small-scale version of the architecture was implemented on FPGA and as ASIC in 28 nm deep sub-micron CMOS technology.

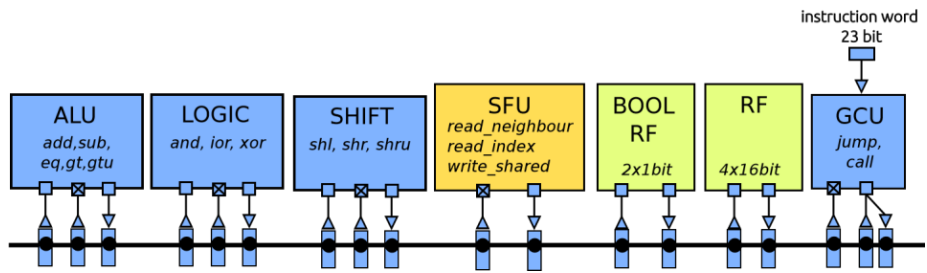

**Fig.4.** Transport Triggered Architecture serves as PE core

**Table 2.** Code examples of the custom FU transports

| Reading Neighbours | 1. 0 → CustomFU.Inp1.read neighbour | 0 for |
| | 2. CustomFU.output → RF.0; | North,... |
| Reading index | 1. 0 → CustomFU.Inp1.read index | 0,1 for |
| | 2. CustomFU.output → RF.0 | X,Y |
| Writing output | 1. RF.0 → Custom FU.Inp1.write Shared | – |
| Data passing | 1. CustomFU.output → CustomFU.Inp1.write Shared | – |

## 4    Results and Discussion

### 4.1    Application Example: Descriptive features

To evaluate the usefulness of our architecture, we have implemented several lowlevel image processing operations, including local descriptors such as the Local Binary Patterns, 3x3 convolutions using integer coefficients and max-pooling. All the operations were tested on the processor. We provide cycle accurate simulations.

**Local descriptors** represent features in small local image patches. Handcrafted local descriptors include binary operators such as Local Binary Patterns (LBP), symmetric operators, e.g., Local Phase Quantization (LPQ), and Binarized Statistical Image Features (BSIF) that are learned from image statistics. They are used in applications that range from face analysis to texture classification [10].

As opposed to handcrafted descriptors, there is a recent surge of learnable local descriptors. This generation of compact and efficient operators is emerging due to schemes that allow individual filters to be learned for different applications and image regions. Examples include regressing Local Binary Features (rLBF) that are utilized in state of the art shape and facial landmark detection [19], or local binary kernels used in neural networks [25] demonstrated in several image classification applications.

The local descriptors share a common computational structure, as they can be expressed in a way that allows for pipelined implementations. The exposed bypass networks of TTA processors enables building these pipelines by software controlled transports.

We decided to evaluate the performance of our processor with the simple, yet useful local descriptors, including LBP [17]. LBP is considered to be computationally cheap, but it needs to be computed for every pixel and is therefore a memory-bound algorithm. In its simplest form, for each pixel value, a binary vector is constructed by comparing the pixel value with values of its immediate surrounding neighbours. Several hardware implementations for efficient LBP extraction have been proposed and thoroughly evaluated [16].

The LBP descriptor can be computed with the proposed processor in a few cycles and using few resources. The number of cycles consumed is 74, while two 16-bit registers and one 1-bit Boolean register suffice. Table 3 contains an excerpt from an LBP TTA transport program. Table 4 summarizes the results of SystemC cycle accurate model simulations for each operation.

**Convolution** is a fundamental image processing operation in which the input is spatially convolved with arbitrary kernels.

We implemented and evaluated a 3x3 convolution on our proposed TTA system utilizing integer and fixed-point calculations. In our implementation, the precise number of consumed cycles depends on the actual kernel weights. Sobel edge detector and Box blur [9] have minimum arithmetic needs and their implementations were evaluated in our experiments. In addition, we implemented

**Table 3.** Code excerpt from LBP program (3 × 3 window)

```
mainloop :
      0 → RF.0;
      0 → RF.3;
      0 → FU.P1.read neighnour;
      FU.P2 → RF.5;
      RF.5 → FU.P1.write Shared;
      RF.6 → alu comp.in2 ;
      5 → FU.P1.read neighnour ;
      FU.P2 → alu comp.in1t.gtu ; alu
      comp.out1 → bool.0 ; 1 →
      RF.1;
      ?bool.0 RF.0 → RF.1; RF.1 →
      alu comp.in2 ; RF.3 → alu
      comp.in1t.add; alu comp.out1
      → RF.3; 6 → FU.P1.read
      neighnour ; FU.P2 → alu
      comp.in1t.gtu ; alu comp.out1
      → bool.0 ; 2 → RF.1;
      ?bool.0 RF.0 → RF.1; RF.1
      → alu comp.in2; RF.3 → alu
      comp.in1t.add; alu comp.out1
      → RF.3;
      ...
```

**Table 4.** Summary of results of SystemC cycle accurate model simulations

| Operations | Window Size | Number of clock cycles |
|---|---|---|
| LBP | 3 × 3 | 74 |
| Conv. (binary weights) | 3 × 3 | 56 |
| Conv. (integer weights) | 3 × 3 | 1553 |
| Max-pooling | 3 × 3 | 271 |

kernels with random weights. The number of cycles required is reported in the simulation results is presented in Table 4.

**Pooling** layers are important in Convolutional Neural Networks [14]. Pooling is a down-sampling operation implemented using a custom stride ( down-sampling factor). Typical CNN architectures commonly use Max-pooling in which the maximum value in a window region is selected.

Pooling operations can be applied on non-overlapping windows. Hence, a method to divide PEs into independent slices is required. Our implementation can achieve this through selecting the PEs with indices that are multiple of our desired stride (e.g in case the stride is equal to two, PEs with indices of multiples of two are activated). To find if an index is multiple of a number, we do not need to compute the remainders since, for example, multiples of 2,3 and 5 can be computed through simple iterative methods [20]. The pooling operation with stride of 2 is depicted in Fig 5. The results for Max-pooling also are presented in Table 4.

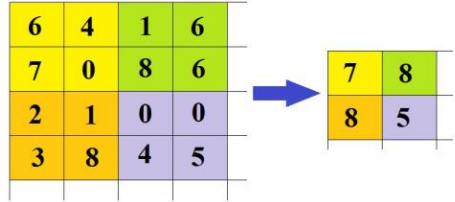

**Fig.5.** Max pooling example with a stride of 2

## 4.2    FPGA and ASIC implementation

To show the implementation feasibility of our processor, and for verification purposes, we have carried out a small-scale design into both FPGA and ASIC. In both implementations we measure and extrapolate estimations of power consumption, occupied area and number of gates.

**FPGA Results:** We carried out our FPGA implementation starting from our SystemC and HDL modelling, and employed an Altera Cyclone IV EP4CE115F29C7 FPGA. Our small scale implementation is comprised of 110 cores ($10 \times 11$). The number of used logic elements and the measured power consumption results are presented in Table 5. The static power is constant for this FPGA regardless of the design and clock frequency. The dynamic power for this $10 \times 11$ array is similar to the static power, and is relatively low for a FPGA implementation.

**Table 5.** Implementation of a $10 \times 11$ TTA array on Cyclone IV FPGA

|  | $10 \times 11$ Array | Single TTA core |
|---|---|---|
| Static Power | 104.30 $mW$ | - |
| Dynamic Power | 113.79 $mW$ | 1 $mW$ |
| Total Power | 234.30 $mW$ | - |
| Logic Cells | 69,983/81,264 (86%) | 630 |
| Clock | 50 MHz | |

**ASIC Results** In addition to the FPGA implementation, we have synthesized our design to an ASIC using 28 $nm$ low power libraries. In addition, we have performed post-layout simulations. We obtained the results of the power estimations per PE (TTA core) based on a small scale implementation. Based on the measurements and simulations, we expect the total power consumption to be roughly proportional to the number of PEs with almost negligible overheads. The following table summarizes the ASIC implementation results for three different settings.

**Table 6.** TTA core ASIC implementation results

| | 0.8 V 25 ℃ | | | 0.6 V 125 ℃ | | | 0.8 V 125 ℃ | | |
|---|---|---|---|---|---|---|---|---|---|
| Clock(MHz) | 1 | 10 | 100 | 1 | 10 | 100 | 1 | 10 | 100 |
| Static Power ($\mu$W) | 0.7 | 0.68 | 0.55 | 8.2 | 8.0 | 8.1 | 15.1 | 15.5 | 15.2 |
| Dynamic Power($\mu$W) | 1.76 | 17.51 | 184 | 0.96 | 10.4 | 101 | 1.8 | 19.1 | 186 |
| Total Power($\mu$W) | 2.4 | 18.1 | 185 | 9.2 | 18.4 | 109 | 16 | 34 | 202 |
| Static / Total Power | 0.29 | 0.04 | 0.003 | 0.89 | 0.44 | 0.07 | 0.89 | 0.45 | 0.07 |
| Area | $55\ \mu m \times 55\ \mu m$ | | | | | | | | |

Each PE occupies an area of 55 $\mu m$ × 55 $\mu m$ while the array size growth is almost linear with the number of PEs. Extrapolating, we can expect that a array of 128 × 128 processors, would occupy around 6.5 $mm$ × 6.5 $mm$. We expect the leakage current to be substantially lower in typical settings, for the near-threshold results.

The available technology libraries allowed us to only carry out simulation in extremes corners for 0.6 V (i.e. 125 ℃ and -40 ℃). Therefore, simulation results for both 0.8 V typical (25 ℃) and 0.8 V worst case (125 ℃) are added to help understanding the impact of leakage to total power consumption. Based on these results static power is expected to have a trivial portion in typical temperature (25 ℃) for operating point of 0.6 V.

Based on the simulation results, the array consumes 18 $uW$ per core at 10 MHz operation. Considering the results in Table 4, the processor is fast enough to complete multiple image operations and to be turned off (clock gated) before the next frame comes available.

## 5    Discussion and future work

In our design, we explored implementing massive array processors with TTA processing elements operating in near-threshold region. Our results appear promising, in particular, when considering the programmable flexibility of the solution, a feature that is not present in similarly power efficient solutions for the same purpose. The energy consumed per pixel is just 1.4 $nJ$ per pixel in the FPGA case for each LBP operation. In case of the ASIC implementation at 10 MHz clock frequency (0.6 V worst case 125 ℃), the energy dissipation is around 0.17 $nJ$ per pixel. Our results are very close to the best ones achieved for hardwired LBP implementations [16].

Future work includes investigation of race-to-sleep schemes [23], which could reduce the average power consumption. Depending on the required image operations and the input frame rate, the array could be turned off for relatively long periods, permitting to tolerate wake-up overheads.